\documentclass[a4paper,11pt]{article}
\pdfoutput=1 % if your are submitting a pdflatex (i.e. if you have
\usepackage{physics} 
\usepackage{jheppub} % for details on the use of the package, please
\usepackage[T1]{fontenc} % if needed
\usepackage{tikz}
\usepackage{braket}
\usepackage{bbm}
\usepackage{comment}
\newtheorem{lemma}{Lemma} 
\newcommand{\lr}[1]{\left( #1 \right)}
\definecolor{Bcolor}{RGB}{0,0,255}
\definecolor{Rcolor}{RGB}{255,0,0}
\definecolor{Gcolor}{RGB}{0,255,0}

\usetikzlibrary{external}
\newcommand{\JKF}[1]{\textcolor{Bcolor}{#1}}

\title{R\'enyi Mutual Information in Quantum Field Theory, Tensor Networks, and Gravity}

\author[1,2]{Jonah Kudler-Flam,}
\affiliation[1]{School of Natural Sciences, Institute for Advanced Study, Princeton, NJ 08540 USA}
\affiliation[2]{Princeton Center for Theoretical Science, Princeton University, Princeton, NJ~08540, USA}
\emailAdd{jkudlerflam@ias.edu}
\author[3,4]{Laimei Nie,}
\emailAdd{nlm@purdue.edu}
\affiliation[3]{Department of Physics and Institute for Condensed Matter Theory, University of Illinois at Urbana-Champaign, Urbana, Illinois 61801, USA}
\author[3]{Akash Vijay}
\affiliation[4]{Department of Physics and Astronomy, Purdue University, West Lafayette, Indiana 47906, USA}
\emailAdd{akashv2@illinois.edu}

\abstract{We explore a large class of correlation measures called the $\alpha-z$ R\'enyi mutual informations (RMIs). Unlike the commonly used notion of RMI involving linear combinations of R\'enyi entropies, the $\alpha-z$ RMIs are positive semi-definite and monotonically decreasing under quantum operations, making them sensible measures of total (quantum and classical) correlations. This follows from their descendance from R\'enyi relative entropies. In addition to upper bounding connected correlation functions between subsystems, we prove the much stronger statement that for certain values of $\alpha$ and $z$, the $\alpha-z$ RMIs also lower bound connected correlation functions. We develop an easily implementable replica trick which enables us to compute the $\alpha-z$ RMIs in a variety of many-body systems including conformal field theories, free fermions, random tensor networks, and holography.}

\begin{document} 
\maketitle
\flushbottom

\section{Introduction}
\iffalse
\JKF{Notation
\begin{itemize}
    \item For relative entropies, use $|| $ between arguments. The arguments should be \textit{density matrices}. The R\'enyi index should \textit{always} be in $\alpha$-z notation e.g. for Sandwiched RRE, $D_{\alpha,\alpha}(\rho||\sigma)$
    \item For mutual informations, use $;$ between arguments. The arguments should be $subregions$. The R\'enyi index should \textit{always} be in $\alpha$-z notation as the subscript e.g. for Petz RMI $I_{\alpha,1}(A;B)$
    \item For R\'enyi entropies, the argument should be one density matrix and the R\'enyi index should be a subscript e.g. $S_n(\rho) = \frac{1}{1-n}\log \Tr\left( \rho^n\right)$. For von Neumann entropy, use $S_{vN}$ instead of $S_1$.
    \item Whereever possible, use parenthesis instead of brackets or curly brackets
    \item For definitions, use $:=$
    \item For twist operators, use $\mathcal{T}$; For permutation, use $g$.
    \item For scaling dimension of twist operators, use $\Delta_n$ instead of $d_n$
\end{itemize}
}
\fi

Quantum information theory has proven to be tremendously valuable in understanding the nature of spatial correlations which are ubiquitous in many-body systems and quantum field theories. Much attention has been devoted to the von Neumann (entanglement) entropy
\begin{align}
    S_{vN}(\rho) := -\Tr (\rho\log \rho)
\end{align}
where $\rho$ is a density matrix i.e.~a positive semi-definite, Hermitian matrix with unit trace. 
If the state of the system happens to be pure, then a non-zero von Neumann entropy for a subregion $A$ can be attributed to entanglement between the degrees of freedom contained in $A$ and its complement. If the state of the system is not pure, then the entropy
is not a faithful measure of correlations. To be a faithful measure of correlations, a quantity must be monotonically decreasing under quantum operations which are completely positive trace-preserving (CPTP) maps. This is sometimes referred to as the data processing inequality (DPI).

To obtain a genuine correlation measure between two subsystems $A$ and $B$, one may take a linear combination of von Neumann entropies, called the mutual information
\begin{align}
\label{Eq:MI_Definition}
    I(A;B) :=  S_{vN}(\rho_{A}) + S_{vN}(\rho_{B}) - S_{vN}(\rho_{AB}).
\end{align}
The non-negativity and fulfillment of DPI of the mutual information are equivalent to the statements that von Neumann entropy is subadditive \cite{araki1970entropy} and strong subadditive \cite{lieb1973proof} respectively.
The mutual information has many additional nice properties, one of the most physically relevant being that it bounds all connected correlation functions \cite{2008PhRvL.100g0502W}. Namely, if $\mathcal{O}_{A}$ and $\mathcal{O}_{B}$ are operators with support on regions $A$ and $B$ respectively, then the connected correlator $\langle\mathcal{O}_{A}\mathcal{O}_{B}\rangle_{c} := |\langle\mathcal{O}_{A}\mathcal{O}_{B}\rangle - \langle\mathcal{O}_{A}\rangle\langle\mathcal{O}_{B}\rangle|$ is upper bounded by the mutual information
\begin{align}
\frac{\langle\mathcal{O}_{A}\mathcal{O}_{B}\rangle_{c}^2}{2\norm{\mathcal{O}_A}_{\infty}\norm{\mathcal{O}_B}_{\infty}} \leq I(A;B).
\end{align}
In the denominator, we have employed the ``operator'' or ``infinty''-norm which is a limit of the Schatten $k$-norms
\begin{align}
\label{eq:Schattenknorm}
    \norm{X}_k := \lr{\Tr \lr{\sqrt{X^{\dagger}X}}^k}^{1/k}.
\end{align}

When proving properties of the mutual information, it is often useful to re-express the mutual information as a relative entropy 
\begin{align}
I(A;B) = D(\rho_{AB}||\rho_{A}\otimes\rho_{B}),
\end{align}
where
\begin{align}
D(\rho||\sigma) := \Tr(\rho \log \rho - \rho \log \sigma).
\end{align}
The relative entropy is an asymmetric measure of distinguishability between states $\rho$ and $\sigma$, possessing key properties including non-negativity, vanishing only when the two states are equal, and fulfillment of the DPI \cite{lindblad1975completely}.
The mutual information, defined as the relative entropy between $\rho_{AB}$ and $\rho_{A}\otimes\rho_{B}$, thus tells us to what extent the state $\rho_{AB}$ does not factorize across systems $A$ and $B$, i.e.~how much they are correlated.

It is natural to inquire if there are other measures of correlation that are both computable and provide complementary information to the mutual information.
A common substitute for the mutual information is the R\'enyi mutual information (RMI) that replaces the von Neumann entropies in \eqref{Eq:MI_Definition} with R\'enyi entropies
\begin{align}
\label{eq:naiveRenyiMI}
    I_{n}(A;B) :=  S_{n}(\rho_{A}) + S_{n}(\rho_{B}) - S_{n}(\rho_{AB}),\quad S_n(\rho):= \frac{1}{1-n}\log \Tr\left(\rho^n\right).
\end{align}
While the RMI mimics the mutual informations fairly well in many situations, it is known that this quantity is somewhat pathological and cannot be interpreted as a measure of correlation because it is neither non-negative nor does it satisfy the DPI in regimes of physical relevance. Said another way, the R\'enyi entropies do not obey either subadditivity %($S_{n}(A)+S_{n}(B) \geq S_{n}(AB)$)
nor strong subadditivity \cite{2013RSPSA.46920737L}. 
One can find simple counter-examples as we demonstrate in Section \ref{sec:simple_example}, where the RMI defined in~\eqref{eq:naiveRenyiMI} fails to be non-negative, which furthermore implies violation of the DPI.

In this paper, we investigate a different R\'enyi generalization of the mutual information inspired from its expression as a relative entropy rather than its expression as a linear combination of von Neumann entropies. In particular, we consider a two-parameter family of R\'enyi divergences, dubbed the $\alpha-z$ R\'enyi relative entropies \cite{2015JMP....56b2202A}
\begin{align}
\label{eq:alphazRRE}
    D_{\alpha,z}(\rho||\sigma) := \frac{1}{\alpha - 1}\log \Tr\bigg(\sigma^{\frac{1-\alpha}{2z}}  \rho^{\frac{\alpha}{z}}  \sigma^{\frac{1-\alpha}{2z}} \bigg)^{z} .
\end{align}
The $\alpha-z$ R\'enyi relative entropies obey the DPI for a large region of the $(\alpha,z)$ plane (see \cite{2015JMP....56b2202A} for details). There are two sub-families that have garnered significant attention. These are the Petz R\'enyi relative entropy (PRRE) \cite{lieb1973convex, uhlmann1977relative, petz1986quasi} and the sandwiched R\'enyi relative entropy (SRRE) \cite{2014CMaPh.331..593W,2013JMP....54l2203M}
\begin{align}
\label{eq:PetzRRE}
    D_{\mbox{\tiny Petz},\alpha}(\rho||\sigma) &:= D_{\alpha, 1}(\rho||\sigma) =  \frac{1}{\alpha - 1}\log \Tr\bigg( \rho^{\alpha}  \sigma^{{1-\alpha}} \bigg) 
\\
\label{eq:SandwichedRRE}
    D_{\mbox{\tiny sand},\alpha}(\rho||\sigma) &:= D_{\alpha,\alpha}(\rho||\sigma) =  \frac{1}{\alpha - 1}\log \Tr \bigg(\sigma^{\frac{1-\alpha}{2\alpha}}  \rho \sigma^{\frac{1-\alpha}{2\alpha}} \bigg)^{\alpha} .
\end{align}
The PRRE obeys the DPI for $\alpha \in [0,2]$ and was analyzed in \cite{2023PhRvL.130b1603K}, 
whereas the SRRE does so for $\alpha \in [1/2,\infty)$. Both limit to the relative entropy as $\alpha$ is taken to one. 
It is clear that we may now define the $\alpha-z$ R\'enyi mutual information as 
\begin{align}
    I_{\alpha,z}(A;B) := D_{\alpha,z}(\rho_{AB}||\rho_A \otimes \rho_B)
\end{align}
By construction, it is non-negative and obeys the DPI. Thus, it is a genuine measure of correlation. 

A special case that will be useful for consistency checks later on is when $\rho_{AB}$ is a pure state, where the following equality holds between $I_{\alpha,z}(A;B)$ and a specific R\'enyi entropy of region $A$
\begin{align}
    I_{\alpha,z}(A;B) = 2S_{\frac{2(1-\alpha)}{z} + 1} (\rho_A)= 2S_{\frac{2(1-\alpha)}{z} + 1} (\rho_B).
\end{align}
To prove this, consider
the following Schmidt decomposition of the pure state $\ket{\psi}$ on $A\cup B$
\begin{align}
    \ket{\psi} &= \sum_{i} \sqrt{p_{i}} \ket{i_{A}}\ket{i_{B}}, \ \ \sum\limits_i p_i = 1 
\end{align}
which yields
\begin{align}
\label{eq:SchmidtrhoAB}
    \rho_{AB} &= \sum_{ij} \sqrt{p_{i}p_{j}} \ket{i_{A}}\ket{i_{B}}\bra{j_{A}}\bra{j_{B}}, \ \ 
    \rho_{A}\otimes \rho_{B} =  \sum_{ij} {p_{i}p_{j}} \ket{i_{A}}\ket{j_{B}}\bra{i_{A}}\bra{j_{B}}.
\end{align}
It is straightforward to check that
\begin{align}
\label{eq:pure_limit}
    I_{\alpha,z}(A;B) = \frac{z}{\alpha-1} \log \Big( \sum\limits_i p_i^{ \frac{2(1-\alpha)}{z}+1} \Big)
\end{align}
which is equal to $ 2S_{\frac{2(1-\alpha)}{z}+1} (\rho_A)$.

The $\alpha-z$ RMIs each provide complementary information. One way to see this is through their operational interpretations via quantum hypothesis testing, where each value of $\alpha$ and $z$ characterizes the extent to which $\rho_{AB}$ and $\rho_A \otimes \rho_B$ may be distinguished using quantum measurements. A novel complementary aspect of these RMIs is that for particular values of $\alpha$ and $z$, they provide \textit{lower} bounds on connected correlation functions.

\paragraph{Organization}

In Section \ref{sec:simple_example}, we provide a simple example where \eqref{eq:naiveRenyiMI} is seen to be negative and thus not meaningful. We demonstrate how the $\alpha-z$ RMIs avoid this pathological behavior.

In Section \ref{sec:ReplicaTrick}, we develop a replica trick for $\alpha-z$ R\'enyi relative entropies and RMIs. This involves considering the joint integer moments of $\rho_{AB}$ and $\rho_A\otimes \rho_B$ and applying an appropriate analytic continuation. For Petz RMI and sandwiched RMI, we show how the replica trick simplifies.

In Section \ref{sec:bounds}, we prove that the $\alpha-z$ RMIs both upper and lower bound connected correlation functions.

In Section \ref{sec:CFT}, we expand upon the previously developed replica trick in 2D conformal field theories by introducing quasi-local twist operators whose correlation functions evaluate the $\alpha-z$ RMIs. We identify their scaling dimensions and OPE coefficients, and compute their values in situations where their behavior is fixed by conformal symmetry and thus theory-independent.

In Section \ref{sec:FreeFermions}, we give expressions for the $\alpha-z$ RMIs for Gaussian states of fermionic theories. These are formulas involving just the two-point function and are efficiently evaluated numerically, with computational cost growing polynomially with system size instead of the standard exponential cost in quantum mechanics. At the critical point (massless), we demonstrate that the ground state RMIs agree with those computed in the previous section.

In Section \ref{sec:RTN}, we consider states on tensor networks where each tensor is drawn randomly from a Gaussian distribution. These networks are good models for certain many-body states including holographic conformal field theories because they obey an area law. We expand upon this by computing the $\alpha-z$ RMIs in both AdS/CFT and in simple models of black hole evaporation i.e.~``Page curves'' for the RMIs.

In Section \ref{sec:discussion}, we conclude with a few loose ends such as the rigorous definitions of RMIs in algebraic quantum field theory, prospects for characterizing multipartite entanglement through defining R\'enyi Markov gaps, and the role of symmetries.

\section{A Simple Example}
\label{sec:simple_example}

In this section, we explicitly show an example where \eqref{eq:naiveRenyiMI} is negative but the $\alpha-z$ RMIs are positive. Consider a Hilbert space of two qudits $\mathcal{H} = \mathcal{H}_A \otimes \mathcal{H}_B$ and the global state\footnote{We adapt this example from \cite{2013RSPSA.46920737L}.}
\begin{align}
    \rho_{AB} = \sum_{i,j = 0}^{d-1} P_{ij} \ket{i}\bra{i}_A \otimes \ket{j}\bra{j}_B
\end{align}
where $P_{ij}$ equals $(1-p)$ if $i = j = 0$, $p/(d-1)^2$ if $i,j\neq 0$, and zero otherwise. This is a classically correlated state. The reduced density matrices are 
\begin{align}
    \rho_A = \sum_{i= 0}^{d-1}  \tilde{P}_{i} \ket{i}\bra{i}_A ,\quad \rho_B =  \sum_{i= 0}^{d-1} \tilde{P}_{i} \ket{i}\bra{i}_B ,\quad  \tilde{P}_i := \sum_j P_{ij}.
\end{align}
It is straightforward to evaluate the R\'enyi entropies
\begin{align}
\begin{aligned}
S_n(\rho_{AB}) &= \frac{1}{1-n} \log \left((1-p)^n + \frac{p^n}{(d-1)^{2(n-1)}} \right)
\\
S_n(\rho_A) &= S_n(\rho_B) = \frac{1}{1-n} \log \left( (1-p)^n+ \frac{p^n}{(d-1)^{n-1}} \right)
\end{aligned}
\end{align}
We can now take
\begin{align}
    p = \left( \frac{e^{S(1-n)}-1}{d^{2(1-n)}}\right)^{1/n}, \quad 0\leq S\leq 2 \log d.
\end{align}
 For large $d$ and $0<n<1$, the R\'enyi mutual information can clearly become very negative
\begin{align}
    I_n(A;B) = -S.
\end{align}
Further simple examples for other values of $n$ can be found in \cite{2013RSPSA.46920737L}.

On the other hand, the $\alpha-z$ RMIs are always positive
\begin{align}
    I_{\alpha,z}(A;B) = \frac{1}{\alpha-1} \log \left( (1-p)^{2-\alpha}+p^{2-\alpha}\right) \geq 0.
\end{align}
Notably, this is independent of both $z$ and $d$.

\section{Replica Trick}
\label{sec:ReplicaTrick}
The replica trick is a powerful analytic technique which enables us to evaluate various entanglement and distinguishability measures. The central idea is that integer (joint) moments of density matrices are frequently within analytic control because they only involve matrix multiplication and traces. Many quantities of interest involve non-integer powers or matrix logarithms. The replica trick computes these as analytic continuations of integer moments. This is particularly powerful in quantum field theory where the integer moments are computable from partition function on ``replica manifolds''.

Suppose that we are presented with a quantum field theory with local Lagrangian $\mathcal{L}[\phi]$. Here $\phi$ collectively denotes all of the ``fundamental fields'' of the theory.
Many states, such as the vacuum, may be prepared by a Euclidean path integral. The integer (say $n$) moments of a density matrix for a subregion of this state is equivalent to the partition function on a Riemannian replica manifold, $\mathcal{M}$. This Riemannian manifold is constructed by replicating the original path integral $n$ times and stitching together these replicas cyclically along the subregion
\begin{align}\label{eq:Replica_trick}
    \Tr(\rho^{n}) = \int [D\phi]_{\mathcal{M}} e^{-\int_{\mathcal{M}}\mathcal{L}[\phi]}  ,
\end{align}
For more detail, we refer the reader to \cite{2009JPhA...42X4005C}.

\subsection{R\'enyi Relative Entropies and Mutual Informations}
\label{subsec:GeneralCase}
\begin{figure}
    \centering
    \includegraphics[width = .8\textwidth]{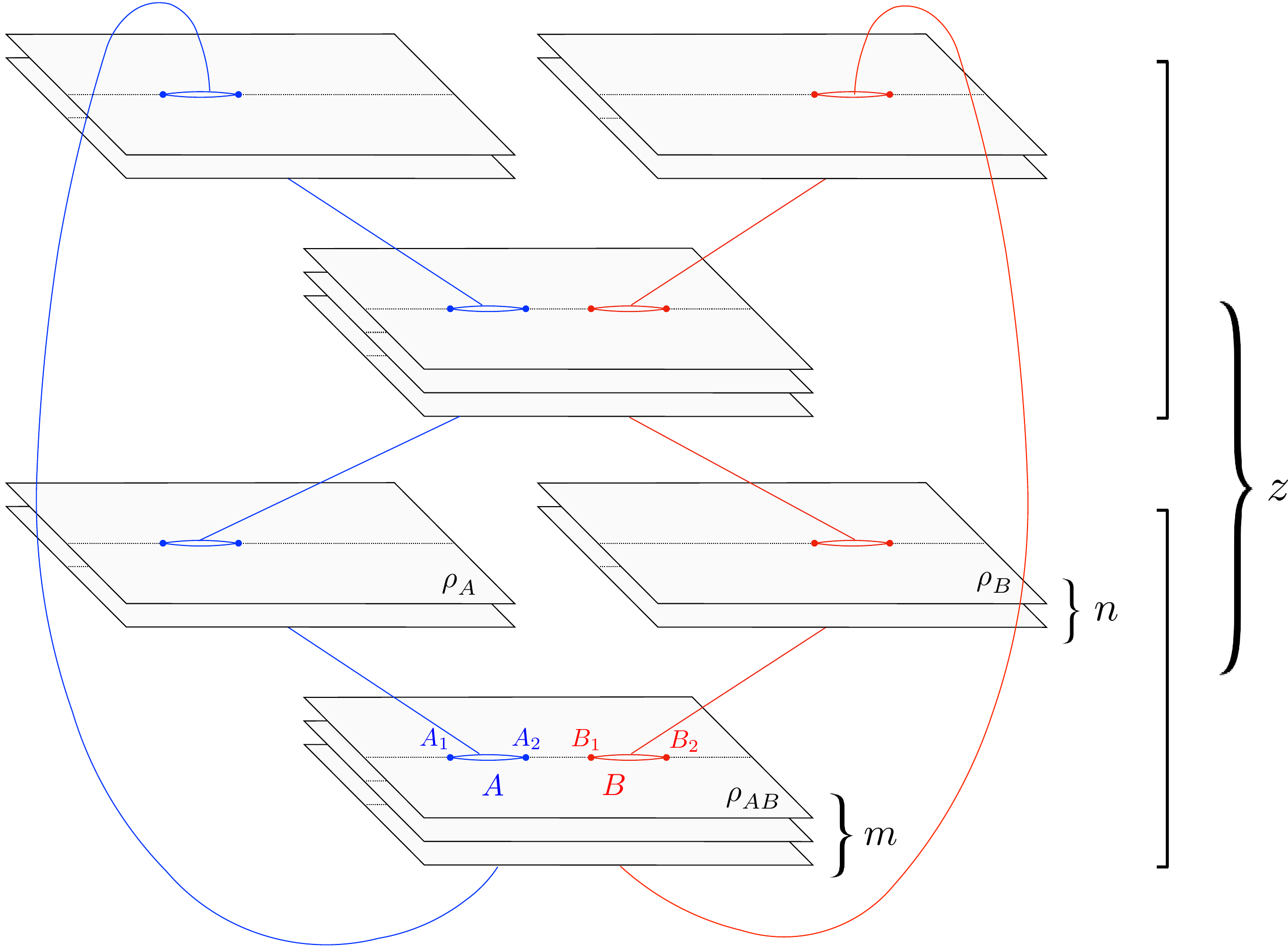}
    \caption{The replica manifold corresponding to $ \Tr \Big( \rho_{AB}^m (\rho_A\otimes \rho_B)^n \Big)^z$ with $m = 3, n = 2, z = 2$ in a $(1+1)d$ system as an example. The solid lines represent identifying the corresponding cuts on neighboring replicas. $A_{1,2}$ and $B_{1,2}$ mark the end points of the intervals $A$ and $B$, respectively.}
\label{fig:alphazReplicaTrick}
\end{figure}

We begin with the most general case of the $\alpha-z$ R\'enyi relative entropy, as defined in~\eqref{eq:alphazRRE}. Using the cyclicity of the trace\footnote{The cyclicity holds even when $z$ is fractional.~\cite{2015JMP....56b2202A}}, these can be rewritten as
\begin{align}
    D_{\alpha,z}(\rho ||\sigma)  = \frac{1}{\alpha - 1} \log \Tr \lr{ \rho^{\frac{\alpha}{z}} \sigma^{\frac{1-\alpha}{z} }}^z.
\end{align}
We then use the following triple replica trick
\begin{align}
\label{eq:alphazRREReplicaTrick}
    D_{\alpha,z}(\rho ||\sigma)  =\frac{1}{\alpha-1} \lim_{m\rightarrow \frac{\alpha}{z}, n\rightarrow \frac{1-\alpha}{z}} \log \Tr(\rho^m \sigma^n)^z,
\end{align}
where $m$, $n$, and $z$ are first taken to be integers, then analytically continued to any real value.
For the $\alpha-z$ R\'enyi mutual information, we use this replica trick for the relevant density matrices
\begin{align}
\label{eq:alphazRMIReplicaTrick}
    {I}_{\alpha,z}(A;B) = \frac{1}{\alpha-1}\lim_{m\rightarrow \frac{\alpha}{z}, n\rightarrow \frac{1-\alpha}{z}} \log \Tr \Big( \rho_{AB}^m (\rho_A\otimes \rho_B)^n    \Big)^z
\end{align}
where the trace term corresponds to the replica manifold structure shown in Fig.~\ref{fig:alphazReplicaTrick}.

The Petz R\'enyi relative entropy, as defined in~\eqref{eq:PetzRRE}, is a simplification using only two replica indices \cite{2021PRXQ....2d0340K}
\begin{align}
     D_{\alpha,1}(\rho ||\sigma)  =\frac{1}{\alpha-1}\lim_{m\rightarrow 1-\alpha} \log \Tr \lr{\rho^{\alpha} \sigma^{m}}.
\end{align}
The Petz RMI is consequently
\begin{align}
    I_{\alpha,1}(A;B) = \frac{1}{\alpha-1}\lim_{m\rightarrow 1-\alpha} \log \Tr \lr{\rho_{AB}^{\alpha} \lr{\rho_A\otimes \rho_B}^{m}}.
\end{align}

The sandwiched R\'enyi relative entropy, as defined in~\eqref{eq:SandwichedRRE}, similarly only requires two replica indices \cite{2014PhRvL.113e1602L,2021PRXQ....2d0340K}
\begin{align}
    D_{\alpha,\alpha}(\rho ||\sigma)  =\frac{1}{\alpha-1} \lim_{m\rightarrow \frac{1-\alpha}{2\alpha}}\log \Tr \lr{\sigma^{m}\rho \sigma^{m}}^{\alpha}
\end{align}
leading to the sandwiched RMI
\begin{align}
    {I}_{\alpha,\alpha}(A;B) = \frac{1}{\alpha-1}\lim_{m\rightarrow \frac{1-\alpha}{2\alpha}} \log \Tr \Big(\lr{\rho_A\otimes \rho_B}^{m}\rho_{AB} \lr{\rho_A\otimes \rho_B}^{m}\Big)^{\alpha}.
\end{align}

\section{Bounds on Correlation Functions}

\label{sec:bounds}

One of the most appealing aspects of applying information theoretic tools to many-body physics is that the information theoretic quantities may be defined and compared between different systems even when the operator content or number of dimensions are very different. Nevertheless, one is frequently interested in specific correlation functions. These may, for example, be useful order parameters for characterizing the phase. In this section, we discuss how mutual information and the various R\'enyi generalizations place strong bounds on all correlation functions in the given quantum state. We first review the classic result for the mutual information from \cite{2008PhRvL.100g0502W}, then prove new constraints.

Consider operators with support in regions $A$ and $B$, $\mathcal{O}_A$  and $\mathcal{O}_B$. The absolute value of their connected correlation function is 
\begin{align}
\langle\mathcal{O}_{A}\mathcal{O}_{B}\rangle_{c} := |\langle\mathcal{O}_{A}\mathcal{O}_{B}\rangle - \langle\mathcal{O}_{A}\rangle\langle\mathcal{O}_{B}\rangle|= |\Tr\left( \rho_{AB} \mathcal{O}_A \mathcal{O}_B\right) - \Tr\left(\left( \rho_{A}\otimes \rho_B\right) \mathcal{O}_A \mathcal{O}_B\right)|.
\end{align}
H\"{o}lder's inequality between Schatten $k$-norms (defined in Eq.~\eqref{eq:Schattenknorm}) states that
\begin{align}
    \norm{XY}_1 \leq \norm{X}_p \norm{Y}_q,\quad p,q \in [1,\infty)
\end{align}
when $p^{-1}+ q^{-1} =1$. Taking $X = \rho_{AB} - \rho_A\otimes\rho_B$ and $Y = \mathcal{O}_A\mathcal{O}_B$, a limit of H\"{o}lder's inequality gives
\begin{align}
\begin{aligned}
    \norm{(\rho_{AB} - \rho_A\otimes\rho_B)\mathcal{O}_A\mathcal{O}_B}_1 &\leq \norm{\rho_{AB} - \rho_A\otimes\rho_B}_1\norm{\mathcal{O}_A\mathcal{O}_B}_\infty
    \\
    &= 2T(\rho_{AB},\rho_A \otimes \rho_B)\norm{\mathcal{O}_A\mathcal{O}_B}_\infty ,
    \end{aligned}
\end{align}
where we have defined the trace distance, $T(\rho,\sigma):=\frac{1}{2}\norm{\rho-\sigma}_1$, which is a metric on the space of density matrices. The left hand side must be at least as large as the absolute value of the connected correlator, so
\begin{align}
    T(\rho_{AB},\rho_A \otimes \rho_B) \geq \frac{\langle\mathcal{O}_{A}\mathcal{O}_{B}\rangle_{c}}{2\norm{\mathcal{O}_A}_{\infty}\norm{\mathcal{O}_B}_\infty }
\end{align}
We may now use Pinsker's inequality
\begin{align}
    D(\rho || \sigma) \geq 2T(\rho ,\sigma)^2,
\end{align}
which leads to 
\begin{align}
\label{eq:vNBound}
    I(A;B) \geq \frac{\langle\mathcal{O}_{A}\mathcal{O}_{B}\rangle_{c}^2}{2\norm{\mathcal{O}_A}_{\infty}^2\norm{\mathcal{O}_B}_\infty^2 }.
\end{align}
Therefore, if the mutual information is small, so are all connected correlation functions. This has the advantage of being operator independent, so that no correlations functions can be overlooked.

The R\'enyi relative entropies satisfy R\'enyi versions of the Pinsker's inequality in the regimes where they obey the data processing inequality \cite{2021arXiv210301709S}
\begin{align}
\label{eq:RenyiPinsker}
   D_{\alpha,1}(\rho || \sigma)\geq{D}_{\alpha,\alpha}(\rho || \sigma) \geq 2\min[\alpha, 1]T(\rho ,\sigma) ^2.
\end{align}
We therefore immediately find that
\begin{align}
\label{eq:RenyiBound}
    I_{\alpha,1}(A;B)\geq {I}_{\alpha,\alpha}(A;B) \geq \min[\alpha, 1]\frac{\langle\mathcal{O}_{A}\mathcal{O}_{B}\rangle_{c}^2}{2\norm{\mathcal{O}_A}_{\infty}^2\norm{\mathcal{O}_B}_\infty^2 }.
\end{align}
When $\alpha>1$, the bound from~\eqref{eq:vNBound} is tighter than the bound from~\eqref{eq:RenyiBound} due to the monotonicity of sandwiched relative entropy with respect to $\alpha$. When $\alpha<1$, the bound~\eqref{eq:RenyiBound} can in principle be tighter. Other types of upper bounds originating from variations of H\"{o}lder's inequality and Pinsker's inequality can be found in Appendix~\ref{app:CorrelationFtnBounds}.

At $\alpha = 1/2$, the R\'enyi relative entropies are related to quantum fidelities. Interestingly, these both lower and upper bound the trace distance \cite{kholevo1972quasiequivalence,1997quant.ph.12042F}
\begin{align}
\begin{aligned}
        {1-e^{-D_{1/2,1}(\rho,\sigma)/2}} \leq{1-e^{-{D}_{1/2,1/2}(\rho,\sigma)/2}} \leq T(\rho,\sigma)\\ \leq \sqrt{1-e^{-D_{1/2,1}(\rho,\sigma)}} \leq \sqrt{1-e^{-{D}_{1/2,1/2}(\rho,\sigma)}}.
\end{aligned}
\label{fvg_eq}
\end{align}
We may then obtain strengthened inequalities on connected correlation functions
\begin{align}
\begin{aligned}
      I_{1/2,1}(A;B) \geq {I}_{1/2,1/2}(A;B )\geq \log \left[ \frac{1}{1-\frac{\langle \mathcal{O}_A  \mathcal{O}_B\rangle_c^2}{4\norm{\mathcal{O}_A}^2_\infty\norm{\mathcal{O}_B}^2_\infty}}\right].
    \label{Ihalf_bound}  
\end{aligned}
\end{align}

More interestingly, we are able to place lower bounds. The norm duality relation implies that
\begin{align}
    \norm{X}_1 = \sup_{\norm{Y}_{\infty} = 1}\Tr (XY)
\end{align}
The operator Schmidt decomposition gives
\begin{align}
    \mathcal{O}_{AB} = \sum_i p_i \mathcal{O}_A^{(i)} \otimes  \mathcal{O}_B^{(i)} 
\end{align}
where $p_i \geq 0$ and the operators in the sum form an orthnormal basis with respect to the Hilbert-Schmidt inner product.
From norm duality, we learn that there exists a unit norm operator such that
\begin{align}
    T(\rho_{AB},\rho_A\otimes \rho_B) = \frac{1}{2}\sum_i p_i \braket{\mathcal{O}_A^{(i)}\mathcal{O}^{(i)}_B}_c
\end{align}
Therefore, using \eqref{fvg_eq}, the RMIs also place \textit{lower bounds} on connected correlation functions
\begin{align}
    {1-e^{-I_{1/2,1}(\rho,\sigma)/2}} \leq{1-e^{-{I}_{1/2,1/2}(\rho,\sigma)/2}} \leq \frac{1}{2}\sum_i p_i \braket{\mathcal{O}_A^{(i)}\mathcal{O}^{(i)}_B}_c.
\end{align}

\section{Conformal Field Theory}

\label{sec:CFT}

We have described in Section~\ref{sec:ReplicaTrick} how one can recast the calculation of various information theoretic quantities into 
the evaluation of a path integral over a replica manifold. 
In the case of $(1+1)d$ conformal field theory, such path integrals may be expressed as correlation functions of quasi-local twist operators, which are primary operators that introduce the appropriate boundary conditions along the cuts in the path integral gluing the different replicas. We briefly summarize the technique and refer the readers to \cite{2009JPhA...42X4005C} for details.

It is convenient to think of the path integral over an $n$-sheeted Riemann surface in terms of a path integral over the single-sheeted Riemann surface used to prepare the state but for the tensor product of $n$ CFTs. The fields on copy $i$ are denoted $\phi^{(i)}$.
We define twist operators $\sigma_{g}$ ($g \in \mathcal{S}_n$, $\mathcal{S}_n$ is the permutation group) which when inserted into an expectation value have the sole effect of creating a branch point in the domain of the path integral. They change the boundary conditions of the path integral by relating the different replica fields in the following way
\begin{align}
    \phi^{(i)}(e^{2\pi i }z + \xi)\sigma_{g}(\xi) = \phi^{(g[i])}(z + \xi)\sigma_{g}(\xi).
\end{align}
The monodromy about the twist operator has implemented the permutation $g$. 

As a simple example, consider the R\'enyi entropy for a single interval $A = [u,v]$ in the ground state. $\Tr(\rho^{n}_{A})$ can then be written as
\begin{align}
    \Tr(\rho^{n}_{A}) \propto \langle {\sigma_{(1\dots n )}}(u) {\sigma_{(n\dots 1 )}}(v) \rangle_{\mathbb{C}} ,
\end{align}
where $(1...n)$ and $(n...1)$ are the standard cyclic notations. This two-point function is entirely fixed by the scaling dimensions of the twist operators, $\Delta_{(1...n)}$ and $\Delta_{(n...1)}$
\begin{align}
    \langle {\sigma_{(1\dots n )}}(u) {\sigma_{(n\dots 1 )}}(v) \rangle_{\mathbb{C}} = \frac{1}{|u-v|^{2\Delta_{(1...n)}}},\quad  \Delta_{(1\dots n )}  =  \Delta_{(n\dots 1 )}= \frac{c}{12}\bigg(n-\frac{1}{n}\bigg),
\end{align}
where $c$ is the central charge. 

For the $\alpha-z$ R\'enyi mutual information, a more complicated set of twist operators are needed to generate the replica manifold in Fig.~\ref{fig:alphazReplicaTrick}:
\begin{align}
\begin{aligned}
     g_A &= \left(\prod_{i = 0}^{z-1} i(m+2n)+1,\dots, i(m+2n)+m+n \right),
     \\ 
     g_B 
     &= \left(\prod_{i = 0}^{z-1} i(m+2n)+1,\dots, i(m+2n)+m,  i(m+2n)+m+n+1,\dots ,(i+1)(m+2n)\right),
     % \\
     % g_{A}^{-1}g_B  &= \prod_{i = 1}^z ().   
\end{aligned}
\end{align}
The products within the parentheses are shorthand notations for the cyclic notations, for example $(\prod_{i=0}^2 3i+1,3i+2,3i+3) = (1,2,3,4,5,6,7,8,9)$.
Below we identify their scaling dimensions and OPE coefficients.

\subsection{Scaling Dimensions}
The scaling dimensions of twist operators that generate the manifold in Fig.~\ref{fig:alphazReplicaTrick} are determined by the cycle structure of the corresponding permutations. In general, the scaling dimension will be
\begin{align}
    \Delta_g  = \sum_{g_i \in \text{cycles}(g)} \frac{c}{12}\left(|g_i|-\frac{1}{|g_i|} \right)    ,
\end{align}
where the absolute value denotes the length of the cycle.

The permutations on $A$ and $B$, $g_A$ and $g_B$, consist of one cycle of length $(m+n)z$ , and $nz$ cycles of length 1. Therefore, the scaling dimension of the corresponding twist operators are
\begin{align}
    \Delta_{{g_A}} = \Delta_{{g_A^{-1}}}= \Delta_{{g_B}} = \Delta_{{g_B^{-1}}}= \frac{c}{12}\left((m+n)z - \frac{1}{(m+n)z} \right)
\end{align}
When $A$ and $B$ are adjacent, the permutation at their intersection point is $g_A^{-1}g_B$. This composite permutation consists of $z$ cycles of length $(2n+1)$ and $(m-1)z$ cycles of length 1. The scaling dimension is then
\begin{align}
    \Delta_{{g_{A}^{-1} g_{B}}}= \frac{zc}{12} \Big(2n+1 - \frac{1}{2n+1}\Big).
\end{align}

\subsection{OPE Coefficients}
\label{subsec:OPEcoefficients}
In the classic work of Lunin and Mathur \cite{2001CMaPh.219..399L}, general formulas are derived for the OPE coefficients of twist fields. Unfortunately, in most cases, these formulas are extremely complicated and there is no clear way to analytically continue them properly for our purposes. Our situation is further complicated by the necessity of twist operators corresponding to multiple non-trivial cycles. Only in the case $z = 1$ (the Petz RMI) are all twist operators single non-trivial cycles. 

When the replica manifold has the topology of a sphere, the OPE coefficients are universal, only depending on the central charge of the theory. We leverage this universality in Section~\ref{sec:FreeFermions} and obtain the OPE coefficients by evaluating the $\alpha-z$ RMI in free fermion CFT. Below we show that the replica manifold's genus is indeed zero in the case of $\alpha-z$ RMI.

Consider an $N$-sheeted surface with $k$ ramification points, with ramification indices $\epsilon_j$ ($j = 1, ..., k$), meaning that $\epsilon_j$ sheets meet at the $j^{th}$ point. The Riemann-Hurwitz formula states that such a surface has genus
\begin{align}
    g = \frac{1}{2}\sum_{j = 1}^k \epsilon_j - \frac{k}{2}-N+1.
\end{align}
We consider the adjacent interval limit of Fig.~\ref{fig:alphazReplicaTrick} ($A_2 = B_1$). At both $A_1$ and $B_2$, there is one ramification point with index $(m+n)z$ and $nz$ with index $1$. At $A_2 = B_1$, there are $z$ ramification points with index $(2n+1)$ and $(m-1)z$ ramifications points with index $1$. Therefore, the genus of the Riemann surface in Fig.~\ref{fig:alphazReplicaTrick} is
\begin{equation}
g_{\mathcal{M}} = \frac{1}{2}\Big[ 2(m+n)z  + 2nz + (2n+1) z + (m-1)z  \Big] - \frac{2+(2n+m)z }{2} - (m+2n)z + 1 = 0
\end{equation}
Thus, the OPE coefficients are completely universal, fixed by the conformal symmetry.

\subsection{Universal Behavior}
We are now ready to deduce universal properties of the $\alpha-z$ RMI in $(1+1)d$ CFTs.

\paragraph{Pure state limit}
We first consider $A$ to be a region of finite length $l_{A}$ and $B = \overline{A}$, the complement of $A$, in the vacuum state.
The $\alpha-z$ RMI is then computed by a two-point function, which is fixed by conformal symmetry 
\begin{align}
\begin{aligned}
        {I}_{\alpha,z}(A;B) &= \frac{1}{\alpha-1}\lim_{m\rightarrow \frac{\alpha}{z}, n\rightarrow \frac{1-\alpha}{z}} \log \braket{\sigma_{g_{A}^{-1} g_{B}}(A_1){\sigma}_{g_{B}^{-1} g_{A}}(A_2)}_{\mathbb{C}} 
    =
    \frac{2 c (z+1-\alpha)}{3 (z+2-2 \alpha )} \log \frac{l_A}{\epsilon} .
\end{aligned}
\end{align}
We have included an ultraviolet regululator $\epsilon$, which is fixed by dimensional analysis and implicit in the definition of the twist operators.
It can be readily seen that the RMI is equal to $2S_{\frac{2(1-\alpha)}{z}+1} (\rho_A)$, in agreement the general result \eqref{eq:pure_limit}. This is a consistency check for our replica trick.

\paragraph{Adjacent Intervals}
We now turn to the case when intervals $A$ and $B$ are adjacent with length $l_{A}$ and $l_{B}$. In this case, the $\alpha-z$ RMI is computed by a 3-point function of twist operators, which is determined by the scaling dimensions up to an OPE coefficient:
\begin{align}
\label{eq:CFTalphaz}
\begin{aligned}
     {I}_{\alpha,z}(A;B) &= \frac{1}{\alpha-1}\lim_{m\rightarrow \frac{\alpha}{z}, n\rightarrow \frac{1-\alpha}{z}} \log \langle \sigma_{g_A}(A_1)\sigma_{g_A^{-1}g_B} (A_2) {\sigma}_{g_B^{-1}}(B_2) \rangle_{\mathbb{C}}
\\
&=  \frac{ c (z+1-\alpha)}{3 (z+2-2 \alpha )} \log \frac{l_A l_B}{\epsilon(l_A + l_B)} + \frac{1}{\alpha-1}\log C_{\alpha,z}.
\end{aligned}
\end{align}
The term involving the OPE coefficient is subleading ($O(1)$) and we do not have an analytic expression for it. Because it is universal, we will be able to numerically evaluate it in the following section using the free fermion CFT. 

Moving away from the ground state, we can put the CFT at finite inverse temperature $\beta$ by preparing the Gibbs state via a path integral on the cylinder. The cylinder is conformally related to the complex plane, so the result for the RMI remains universal
\begin{align}
\begin{aligned}
     {I}_{\alpha,z}(A;B) &= \frac{1}{\alpha-1}\lim_{m\rightarrow \frac{\alpha}{z}, n\rightarrow \frac{1-\alpha}{z}} \log \langle \sigma_{g_A}(A_1)\sigma_{g_A^{-1}g_B} (A_2) {\sigma}_{g_B^{-1}}(B_2) \rangle_{S_1^{\beta}\times \mathbb{R}}
\\
&=  \frac{ c (z+1-\alpha)}{3 (z+2-2 \alpha )} \log \left(\frac{\beta}{2\pi \epsilon }\tanh \frac{\pi l}{\beta}\right) + \frac{1}{\alpha-1}\log C_{\alpha,z},
\end{aligned}
\end{align}
where for simplicity, we have set $l_A = l_B = l$. A key feature of this formula is that for $l \gg \beta$, it does not grow significantly with $l$, indicating area law correlations at finite temperature.

\iffalse
\paragraph{Close disjoint intervals}
Lets now consider the ($\alpha,z$)-RMI for two disjoint intervals. In a CFT, this would be evaluated by a 4-point function of twist fields, which is not fixed by the conformal Ward identities and will generically depend on the full spectrum of primary operators. Nevertheless, the form of the 4-point function has universal asymptotics when the two intervals are very close to or very far from one another. In either case, a single channel dominates the OPE of the twists. In the case of close disjoint intervals, the twists $\tilde{\mathcal{T}}_{g_{A}}$ and $\mathcal{T}_{g_{B}}$ inserted at the near ends of $A$ and $B$ fuse to produce $\mathcal{T}_{g_{A}^{-1}g_{B}}$: \begin{align}
    \tilde{\mathcal{T}}_{g_{A}} \cross \mathcal{T}_{g_{B}} = \mathcal{T}_{g_{A}^{-1}g_{B}} + \cdots
\end{align}
Thus the 4-point function should approach the universal 3-point function. 
\fi

\section{Free Fermions}
\label{sec:FreeFermions}

We now consider fermionic theories with Hamiltonians that are quadratic in the fermionic operators $\{\psi_i ,\psi_j\} = \delta_{ij}$
\begin{align}
    H = -\sum_{i,j}^N h_{ij} \psi_i^{\dagger} \psi_j.
\end{align}
The eigenstates of this class of Hamiltonians are so-called Gaussian states, completely fixed by the two-point function \cite{2003JPhA...36L.205P}
\begin{align}
\label{eq:twopoint}
    C_{ij}^{(\Psi)} := \bra{\Psi}\psi_i^{\dagger}\psi_j\ket{\Psi}.
\end{align}
This presents a dramatic simplification of the reduced density matrices for $N_A$ fermions in subsystem $A$, which are characterized by $N_A\times N_A$ correlation matrices that are submatrices of \eqref{eq:twopoint}. Recall, that a density matrix of $A$ is more generally characterized by a $2^{N_A}\times 2^{N_A}$ matrix.

Consider a correlation matrix for $A \cup B$
\begin{align}
   C = \begin{pmatrix}
    C_{AA} & C_{AB}
    \\
    C_{BA} & C_{BB}
    \end{pmatrix}.
\end{align}
The off-diagonal terms correspond to the correlations between $A$ and $B$. 
The state $\rho_A \otimes \rho_B$ is characterized by the following correlation matrix
\begin{align}
    C' = \begin{pmatrix}
    C_{AA} & 0
    \\
    0 & C_{BB}
    \end{pmatrix}.
\end{align}

The $\alpha-z$ RMIs are functions of $C$ and $C'$. One may generalize the derivations from \cite{2014PhRvE..89b2102B, 2018JHEP...09..166C} to obtain
\begin{align}\label{ff_alphaz}
\begin{aligned}
      {I}_{\alpha,z}(A;B) &= -\frac{\alpha\Tr \log \left(1-C \right)}{1-\alpha}-\Tr \log \left(1-C' \right)
  \\
  &-\frac{\Tr \log \left( 1+ \left(\left(\frac{C}{1-C}\right)^{\frac{1-\alpha}{2z}}\lr{\frac{C'}{1-C'}}^{\frac{\alpha}{z}}{}\left(\frac{C}{1-C}\right)^{\frac{1-\alpha}{2z}}\right)^{z} \right)}{1-\alpha}.
\end{aligned}
\end{align}
The special cases of Petz RMI and sandwiched RMI are immediate.

\begin{figure}
    \centering
    \includegraphics[width = 0.48\textwidth]{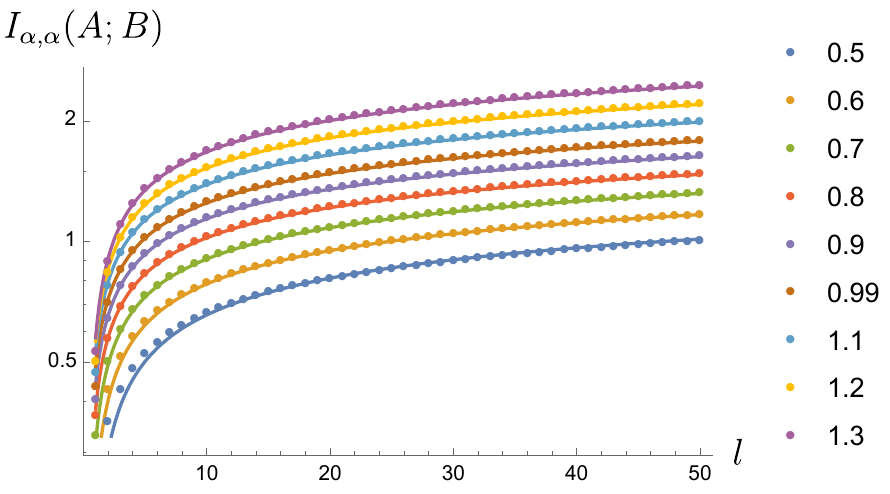}
    \includegraphics[width = 0.48\textwidth]{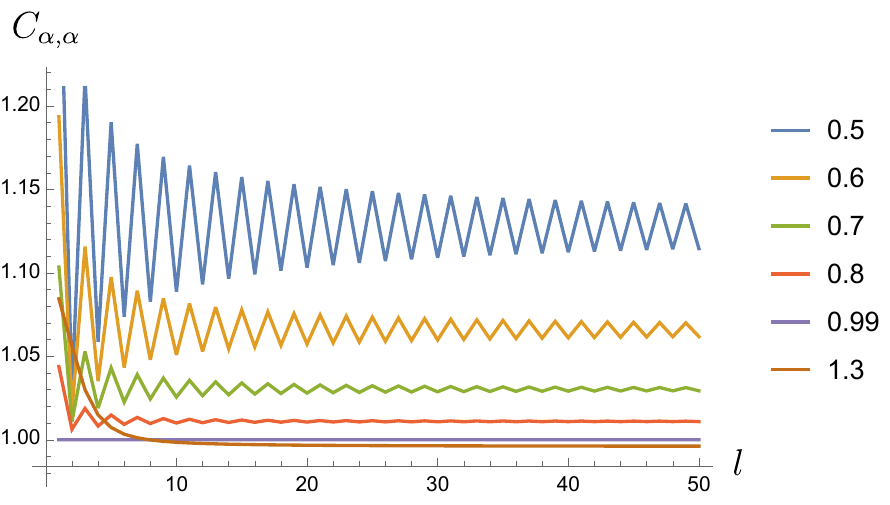}
    \caption{Left: $\alpha-z$ RMI with $l = l_A = l_B$. Different colors represent different values of $\alpha$. The dots are data points while the solid lines are \eqref{eq:CFTalphaz} with the additive constant involving the UV cutoff and OPE coefficient numerically fitted. Right: The difference between the $\alpha-z$ RMI and \eqref{eq:naiveRenyiMI}, which converges to the OPE coefficient at large $l$. The oscillations are an artifact of the RMI from \eqref{eq:naiveRenyiMI} being non-monotonic. For simplicity, we have only plotted the sandwiched RMI ($\alpha = z$) in both plots.}
    \label{fig:sandwiched_MI_free_fermion}
\end{figure}

With \eqref{ff_alphaz} in hand, we may numerically check the CFT formulas from Section \ref{sec:CFT}. We consider a massless free fermion ($c = 1/2$) with discretized Hamiltonian 
\begin{align}
    H = -\frac{i}{2}\sum_{j}\left(\psi^{\dagger}_j\psi_{j+1}-\psi^{\dagger}_{j+1}\psi_{j} \right).
\end{align}
In the ground state, the two-point function is
\begin{align}
    C_{jl} = \begin{cases}
    \frac{(-1)^{j-l}-1}{2\pi i (j-l)} ,  & j \neq l
    \\
    \frac{1}{2}, & j = l
    \end{cases}.
\end{align}
We demonstrate the agreement with CFT results in Figure \ref{fig:sandwiched_MI_free_fermion}.

We furthermore isolate the OPE coefficient by subtracting the $\frac{2-2\alpha+z}{z}$ RMI (namely, $n=\frac{2-2\alpha+z}{z}$ in \eqref{eq:naiveRenyiMI}). The reason we choose this particular RMI is that it has the same UV regulator.\footnote{This method will be explored in detail in \cite{orbifold_OPE} with applications to the AdS$_3$/CFT$_2$ correspondence.} The R\'enyi entropies needed are also expressible in terms of the correlation matrix, leading to 
\begin{align}
    I_\alpha(A; B) = \frac{1}{1-\alpha} \Tr \log   \left( \left(1-C'\right)^\alpha+C'^\alpha\right)-\frac{1}{1-\alpha} \Tr\log   \left( \left(1-C\right)^\alpha+C^\alpha\right).
\end{align}
These are non-monotonic even in the ground state. 

\section{Random Tensor Networks and AdS/CFT}

\label{sec:RTN}

In this section, we study the RMI in tensor networks and in gravity. Tensor networks have proven to be an exceptionally useful framework in many-body and high-energy physics both numerically and theoretically. We focus on a solvable class of tensor networks where each tensor is drawn randomly from an ensemble \cite{2010JPhA...43A5303C, 2016JHEP...11..009H}. Our main focus will be in applications in quantum gravity where they are known to serve as great models of the AdS/CFT correspondence. Moreover, they are directly related to a simple model of black hole evaporation in Jackiw-Teitelboim gravity. 

\subsection{Single Tensor}

We begin with the simplest tensor network consisting of a single random tensor. This may similarly be considered a random state, whose entropy was first studied by Page \cite{1993PhRvL..71.1291P}. We take the Hilbert space to be tripartitioned as $\mathcal{H} = \mathcal{H}_A \otimes \mathcal{H}_B \otimes \mathcal{H}_C$ such that the unnormalized state can be expanded as
\begin{align}
    \ket{\psi} = \sum_{ijk} T_{ijk}\ket{i}_A \ket{j}_B\ket{k}_C,
\end{align}
where the states are orthonormal bases for the subsystems and $T_{ijk}$ are independent Gaussian random variables. Due to the Gaussian behavior, we may evaluate ensemble averages using Wick contractions. We will focus on the Petz RMI, where the the relevant moments are 
\begin{align}
    {\Tr \Big( \rho_{AB}^{\alpha} (\rho_A \otimes \rho_B )^m \Big)} = \frac{\sum_{\tau \in \mathcal{S}_{\alpha + 2m}} d_A^{C(g_A^{-1} \tau)}d_B^{C(g_B^{-1} \tau)}d_C^{C(\tau)}}{\sum_{\tau \in \mathcal{S}_{\alpha + 2m}} (d_Ad_Bd_C)^{C(\tau)}},
\end{align}
where $d_{*}$ is the dimension of Hilbert space $\mathcal{H}_*$, $C(\cdot)$ is the number of cycles in the permutation, and the ensemble averaging is implied. We will always consider the limit of large Hilbert space dimensions, so the denominator will localize, maximized by the identity permutation consisting of $(\alpha + 2m)$ cycles.

When one of the subsystems is much larger than the rest (e.g. $d_A \gg d_B d_C$), then only a single permuation dominates the sum, namely $\tau = g_A$. The truncation of the sum to this single term is
\begin{align}
    \Tr \Big( \rho_{AB}^{\alpha} (\rho_A \otimes \rho_B )^m  \Big)= d_B^{ -2m}d_C^{1-\alpha -m}
\end{align}
and the replica limit gives
\begin{align}
    I_{\alpha,1}(A;B) = \log d_B.
\end{align}
A similar analysis occurs in the  $d_B \gg d_A d_C$ regime such that $ I_{\alpha,1}(A;B)  = \log d_A$. This represents the maximal amount of entanglement. 

When $d_C \gg d_A d_B$, the identity element dominates the sum 
\begin{align}
    \Tr\rho_{AB}^{\alpha} (\rho_A \otimes \rho_B )^m = (d_Ad_B)^{1-\alpha -m}
\end{align}
and the replica limit leads to trivial Petz RMI. This makes sense because both $A$ and $B$ are maximally entangled with $C$. Because entanglement is monogamous \cite{2000PhRvA..61e2306C,2006PhRvL..96v0503O} $A$ and $B$ cannot be entangled with each other.

All results so far for the Petz RMI are in agreement with the naive RMI.
A more interesting regime is when $d_A^2 \sim d_B^2 \sim d_C$. Here, the dominant contributions to the sum come from non-crossing permutations in the first $\alpha$ indices which are enumerated by the Narayana numbers \cite{KREWERAS1972333, SIMION2000367},
\begin{align}
    \Tr\rho_{AB}^{\alpha} (\rho_A \otimes \rho_B )^m = \sum_{ k = 1}^{\alpha} N_{\alpha,k} (d_Ad_B)^{1-m-k}d_C^{k-\alpha} ,\quad N_{\alpha,k} := \frac{1}{\alpha}\binom{\alpha}{k}\binom{\alpha}{k-1}.
\end{align}
It is trivial to analytically continue the expression in $m$. The sum may subsequently be expressed as a hypergeometric function
\begin{align}
    \Tr\rho_{AB}^{\alpha} (\rho_A \otimes \rho_B )^{1-\alpha} = \sum_{ k = 1}^{\alpha} N_{\alpha,k} \lr{\frac{d_A d_B}{d_C}}^{\alpha - k} =\lr{\frac{d_A d_B}{d_C}}^{\alpha-1} \, _2F_1\left(1-\alpha ,-\alpha
   ;2;\frac{d_C}{d_A d_B}\right).
\end{align}
The Petz RMI is therefore
\begin{align}
\label{PRMI_rtn}
    I_{\alpha,1}(A;B) =\begin{cases} \log \lr{\frac{d_A d_B}{d_C}} + \frac{\log \, _2F_1\left(1-\alpha ,-\alpha
   ;2;\frac{d_C}{d_A d_B}\right)}{\alpha-1}, & d_A d_B > d_C
    \\
    \log \lr{\frac{d_C}{d_A d_B}} +\frac{\log \, _2F_1\left(1-\alpha ,-\alpha
   ;2;\frac{d_A d_B}{d_C}\right)}{\alpha-1}, & d_A d_B < d_C
    \end{cases}
\end{align}
This agrees with (a linear combination) of Page's formula when $\alpha$ is taken to be one. When $d_A^2 \sim d_B^2 \gg d_C$, many of the non-crossing permutations become subdominant so that the second term in \eqref{PRMI_rtn} disappears.

In the regime where $d_A d_B = d_C$, the standard RMI is given by
\begin{align}
    I_{3-2\alpha}(A;B) = \frac{1}{2(1-\alpha)}\log{}_2F_1(2(\alpha-1),2\alpha-3;2;1). 
\end{align}
See e.g.~\cite{2021PhRvL.126q1603K} for computations of these R\'enyi entropies. 
This is monotonically decreasing in $\alpha$, the opposite behavior of the Petz RMI and we conclude it does not contain information about the correlations.

\begin{figure}
    \centering
    % \tikzsetnextfilename{phase_tikz}
    % \include{phase_tikz} 
    \includegraphics[width = .5\textwidth]{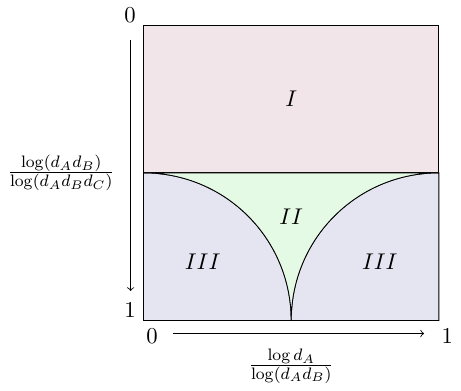}
    \caption{The phase diagram for the Petz RMI for the single tensor network. In the language of \cite{2021PRXQ....2c0347S}, phase I is unentangled, phase II is entanglement saturation, and phase III is maximally entangled.}
    \label{phase}
\end{figure}

 The associated phase diagram is shown in Figure \ref{phase} and is identical to the phase diagram for logarithmic negativity for random states \cite{2021PRXQ....2c0347S}, though distinct from that for reflected entropy \cite{2022JHEP...05..162A}.

\subsection{General Tensor Networks}

So far, we have considered tensor networks without any locality, Haar distributed over the space of quantum states. It is natural to consider tensor networks with many tensors, contracted in nontrivial ways. If all of the tensors are independently random, the ensemeble average over each one will give a sum over the permutation group. In total, the moments are given by a partition function of a Ising-like model with spins valued in $\mathcal{S}_{\alpha + 2m}$
\begin{align}
\begin{aligned}
    \Tr \Big( \rho_{AB}^{\alpha} (\rho_A \otimes \rho_B )^{1-\alpha} \Big) &= {\sum_{\{ g_x\}} e^{-\mathcal{A}[\{g_x \}]}},\\     \mathcal{A}[\{g_x \}] &= \sum_{\{ xy\} \in E}\left(\alpha + 2 m - C(g_x^{-1}g_y)\right)\log d,    
\end{aligned}
\end{align}
where $d$ is the bond dimension of each link, $\{g_x\}$ labels the spin configuration at each tensor, and $E$ is the set of edges in the network. Critically, the model has nontrivial boundary conditions, corresponding to the spins $g_A$, $g_B$, and $e$ (the identity) in regions $A$, $B$, and $C$.
In general, it is very difficult to evaluate such a partition function. The first simplification is that for large $d$, the spin model is in the ferromagnetic phase because the role of inverse temperature is played by $\log d$. This means that the partition function is well approximated by the free energy of domain walls in the bulk of the tensor network, separating the differently aligned spin configurations.

There are further complications when the lengths of the domain walls are degenerate, leading to highly nontrivial combinatorics \cite{2010JPhA...43A5303C,2021PRXQ....2d0340K,2022JHEP...02..076K}. We will assume for simplicity the non-degeneracy of domain wall configurations, in which case, the partition function reduces to the computations in the single-tensor network previously derived. The only modification is that the dimensions of the Hilbert spaces are replaced by the dimensions of the Hilbert spaces on the associated domain walls, $d_{\gamma_*}$. Because we have taken all bond dimensions to be equal, we always land in the phase corresponding to \eqref{PRMI_rtn}. The Petz RMI is
\begin{align}
    I_{\alpha,1}(A;B) =\begin{cases} \log \lr{\frac{d_{\gamma_A} d_{\gamma_B}}{d_{\gamma_C}}} + \frac{\log \, _2F_1\left(1-\alpha ,-\alpha
   ;2;\frac{d_{\gamma_C}}{d_{\gamma_A} d_{\gamma_B}}\right)}{\alpha-1}, & d_{\gamma_A} d_{\gamma_B} > d_{\gamma_C}
    \\
    \log \lr{\frac{d_{\gamma_C}}{d_{\gamma_A} d_{\gamma_B}}} +\frac{\log \, _2F_1\left(1-\alpha ,-\alpha
   ;2;\frac{d_{\gamma_A} d_{\gamma_B}}{d_{\gamma_C}}\right)}{\alpha-1}, & d_{\gamma_A} d_{\gamma_B} < d_{\gamma_C}
    \end{cases}.
\end{align}

\subsection{AdS/CFT and Replica Wormholes}

A central motivation for studying the correlation structure of random tensor networks is due to their close connection with quantum theories of gravity. The two connections that we discuss in this section are so-called fixed-area states \cite{2019JHEP...10..240D,2019JHEP...05..052A} in the AdS/CFT correspondence and replica wormholes in the so-called West Coast model of black hole evaporation \cite{2019arXiv191111977P}.\footnote{To be more specific, we consider the black hole to be in the microcanonical ensemble.} As it turns out, computing R\'enyi mutual information in these states is identical to the single random tensor. The dictionary between the models is that the dimension $d_{*}$ is replaced by $e^{A_{*}/4G}$ in AdS/CFT where $A_*$ is the area of the associated Ryu-Takayanagi surface and $d_{*}$ is replaced by $e^{S_{BH}}$ (black hole entropy) or $k_*$ (number of flavors of end-of-world branes) in the West Coast model. We keep this discussion brief, referring the interested reader to the original literature, because we do not have anything particularly noteworthy to add to this discussion besides simply computing the ``Page curve'' for Petz RMI.

\section{Discussion}

\label{sec:discussion}

We conclude with a few thoughts.

\subsection{Modular Theory}

The elephant in the room is that reduced density matrices and traces for subregions do not exist in quantum field theory. This is due to the universal infinite entanglement in the vacuum state. More formally, it is the statement that the algebras of observables associated to local regions being von Neumann algebras of Type III$_1$ (see e.g.~\cite{2018arXiv180304993W} for a review). Nevertheless, the mutual information for nonadjacent regions is ultraviolet finite, suggesting that such a quantity may be well-defined even in the continuum. This is indeed the case because the relative entropy is well-defined in the continuum and, as we know well, the mutual information is a specific relative entropy. To formalize this, we need to introduce aspects of Tomita and Takesaki's modular theory \cite{tomita1967canonical, takesaki2006tomita}. 

Let $\mathcal{A}$ be a von Neumann algebra\footnote{A von Neumann algebra is a subalgebra of all bounded operators on a Hilbert space, $\mathcal{B}(\mathcal{H})$, that is its own double commutant. The commutant of an algebra is the set of bounded operators that commute with the algebra.} and two states (i.e.~positive maps from $\mathcal{A}$ to $\mathbb{C}$) $\omega$ and $\phi$ with vector representatives
$\ket{\Omega}$ and $\ket{\Phi}$ in a Hilbert space $\mathcal{H}$.
We define the relative Tomita operator as the closure of 
\begin{align}
    S_{\Omega,\Phi; \mathcal{A}} : a\ket{\Phi} \mapsto a^{\dagger} \ket{\Omega}.
\end{align}
The relative modular operator is
\begin{align}
    \Delta_{\Omega,\Phi; \mathcal{A}} := S^{\dagger}_{\Omega,\Phi; \mathcal{A}}S_{\Omega,\Phi; \mathcal{A}}.
\end{align}
In finite dimensions, if $\omega_A$ and $\phi_A$ are the reduced density matrices for $\ket{\Omega}$ and $\ket{\Phi}$ with respect to $\mathcal{A}$, then the relative modular operator is simply $\omega_A \otimes \phi_A^{-1}$. Araki's relative entropy is defined as \cite{araki1976relative}
\begin{align}
    D(\omega,\phi; \mathcal{A}) = - \bra{\Phi} \log \Delta_{\Omega,\Phi; \mathcal{A}} \ket{\Phi}.
\end{align}
While obscured in the present formula, this is independent of the vector representative.
In this language, the mutual information is defined as
\begin{align}
    I(A;B) := D(\omega_{AB}, \omega_A \otimes \omega_B;\mathcal{A}_A\otimes \mathcal{A}_B)
\end{align}
The availability of a tensor product structure is guaranteed when $A$ and $B$ are non-adjacent for theories obeying the split property \cite{doplicher1984standard}. We may then define the Petz RMI and sandwiched RMI in the obvious ways
\begin{align}
\begin{aligned}
        I_{\alpha,1}(A;B) := D_{\alpha,1}(\omega_{AB}, \omega_A \otimes \omega_B;\mathcal{A}_A\otimes \mathcal{A}_B), \\ I_{\alpha,\alpha}(A;B) := D_{\alpha,\alpha}(\omega_{AB}, \omega_A \otimes \omega_B;\mathcal{A}_A\otimes \mathcal{A}_B),
\end{aligned}
\end{align}
where
\cite{2018AnHP...19.1843B,2019JHEP...01..059L}
\begin{align}
\begin{aligned}
    D_{\alpha,1}(\omega,\phi; \mathcal{A}) := \frac{1}{\alpha-1}\log \bra{\Phi} \Delta_{\Omega,\Phi; \mathcal{A}}^{\alpha} \ket{\Phi}
    \\
    D_{\alpha,\alpha}(\omega,\phi; \mathcal{A}) := \sup_{\Psi \in \mathcal{H}} \frac{1}{\alpha-1}\log \bra{\Phi} \Delta_{\Omega,\Psi; \mathcal{A}}^{\alpha} \ket{\Phi}.
    \end{aligned}
\end{align}

\subsection{R\'enyi Markov Gaps}

The mutual information is a measure of correlations between just two subsystems. This is the simplest, and most commonly discussed, form of entanglement. However, multipartite correlations play a crucial role in many important phenomena, for example the physics of thermalization. Unfortunately, the detection and quantification of multipartite entanglement has been notoriously difficult, especially in large systems. 

Some exciting progress came with the discovery of the so-called reflected entropy \cite{2019arXiv190500577D}, which is the von Neumann entropy of a particular purification that represents a mixed state $\rho_{AB}$ as a vector in the natural cone of the GNS Hilbert space $\ket{\sqrt{\rho_{AB}}}$. While the reflected entropy is not itself a bipartite correlation measure \cite{2023PhRvA.107e0401H}, it has been proven to be a useful probe of tripartite entanglement \cite{2020JHEP...04..208A,2021PhRvL.126l0501Z}. 

To understand this, we define the set of pure states called the sum of triangle states (SOTS) on the tripartite Hilbert space $\mathcal{H}_A \otimes \mathcal{H}_B\otimes \mathcal{H}_C$ that only contain pairwise entanglement
\begin{align}
    \ket{\psi }_{ABC} = \bigoplus_{j} \sqrt{p_j}\ket{\psi_j}_{A_R^jB_L^j}\ket{\psi_j}_{B_R^jC_L^j}\ket{\psi_j}_{C_R^jA_L^j}.
\end{align}
Using a structure theorem for states that saturate the strong subadditivity inequality \cite{2004CMaPh.246..359H}, it can proven that a state is a triangle state if and only if \cite{2021PhRvL.126l0501Z}
\begin{align}
    S_R(A;B) = I(A;B),
\end{align}
where $S_R(A;B)$ is the reflected entropy. For all states, $h(A;B) = S_R(A;B)-I(A;B) \geq 0$, so it is reasonable to think of $h(A;B)$, which has been referred to as the Markov gap \cite{2021JHEP...10..047H}, as a measure of tripartite entanglement. Indeed, using the continuity of mutual information and reflected entropy, one can prove that it places a lower bound on the trace distance of the state to the SOTS.

It is natural to consider R\'enyi versions of this quantity. For this, let us express the reflected entropy as a relative entropy
\begin{align}
    S_R(A;B) = D(\rho_{AA^*B}||\rho_{AA^*}\otimes \rho_{B}) ,
\end{align}
where $*$ denotes the dual space of the original Hilbert space. Considering the partial trace over $A^*$, monotonicity of relative entropy tells us that the reflected entropy is larger than the mutual information. We may now define the R\'enyi reflected entropy as\footnote{Note that this is distinct from R\'enyi versions of reflected entropy that have been studied in the literature.}
\begin{align}
    S_R^{\alpha,z}(A;B) = D_{\alpha,z}(\rho_{AA^*B}||\rho_{AA^*}\otimes \rho_{B}).
\end{align}
By monotonicity of the $\alpha-z$ relative entropies, we can define a set of R\'enyi Markov gaps, each of which are positive semidefinite
\begin{align}
    h^{\alpha,z}(A;B)=S_R^{\alpha,z}(A;B)-I_{\alpha,z}(A;B)
\end{align}
It is known that the monotonicity of relative entropy is saturated if and only if all (monotonic) $\alpha-z$ relative entropies saturate, so we arrive at the following lemma
\begin{lemma}
    $h^{\alpha,z}(A;B) = 0$ if and only if $\ket{\psi}_{ABC} \in$ SOTS.
\end{lemma}
It may be interesting to study this family of tripartite entanglement quantities in more depth.

\subsection{Symmetry Resolution}
There has recently been significant attention directed to the symmetry resolution of various entanglement measures, stimulated by the work of Goldstein and Sela \cite{2018PhRvL.120t0602G}. When a theory has a global symmetry, the Hilbert space will fracture as a direct sum over superselection sectors
\begin{align}
    \mathcal{H} = \bigoplus_{q}\mathcal{H}_q.
\end{align}
Density matrices in such systems are also direct sums
\begin{align}
    \rho = \bigoplus_q p_q \rho(q), \quad \sum_q p_q = 1.
\end{align}
It is natural to ask to what extent each symmetry sector contributes to the entanglement in $\rho$. It is straightforward to see that for the von Neumann entropy, 
\begin{align}
    S_{vN}(\rho) = \sum_q p_q S_{vN}(\rho(q)) -\sum_q p_q \log p_q.
\end{align}
The first term is the weighted average of the von Neumann entropies of each sector, individually called the symmetry resolved entanglement entropy, and the second term is the Shannon entropy corresponding to the classical distribution into sectors.

Of course, nothing is stopping us from repeating the very same logic for symmetry resolved R\'enyi mutual information, where we consider R\'enyi relative entropies of the following two density matrices
\begin{align}
    \rho_{AB} = \bigoplus_q p_q \rho_{AB}(q), \quad \rho_{A}\otimes \rho_B  = \bigoplus_{q} \sum_{q_1+q_2 = q}p_{q_1}p_{q_2}\rho_A(q_1)\otimes \rho_B(q_2).
\end{align}
The symmetry resolve relative entropy itself has been studied in \cite{2021JHEP...10..195C} and we expect similar techniques, including in CFT, can be used here.

\acknowledgments
We thank Viktor Eisler, Thomas Faulkner, and Nima Lashkari for helpful discussions and comments. We are particularly grateful to Viktor Eisler for discussions and for sharing his Mathematica code with us. JKF is supported by the Institute for Advanced Study and the National Science Foundation under Grant No. PHY-2207584. LN was partially supported by National Science Foundation under Grant No. DMR-1725401 and the Quantum Leap Challenge Institute for Hybrid Quantum Architectures and Networks under Grant No. OMA-2016136 at the University of Illinois.

\appendix

\section{Another upper bound on connected correlation functions}
\label{app:CorrelationFtnBounds}
Besides the inequalities~\eqref{eq:vNBound} and~\eqref{eq:RenyiBound}, we point out here another upper bound of the two-point connected correlation function.
The R\'enyi version of the Pinsker's inequality~\eqref{eq:RenyiPinsker} involves the trace distance, which is proportional to the Schatten 1-norm. It turns out that when $0< \alpha \le 1$ the Petz RMI is also bounded from below by the Schatten 2-norm, also known as the Hilbert-Schmidt norm \cite{2022arXiv220301964A}
\begin{align}
I_{\alpha, 1} (A;B) \ge \frac{\alpha}{2} ||\rho_{AB} - \rho_A \otimes \rho_B||_2^4
\end{align}
This, combined with the H\"{o}lder inequality
\begin{align}
\label{eq:Holder2}
    \norm{X}_2 \geq \frac{\norm{XY}_1}{\norm{Y}_2 }
\end{align}
yields
\begin{align}
\label{eq:RenyiBoundSchatten2}
    I_{\alpha, 1} (A;B) \ge \frac{\alpha \langle \mathcal{O}_A \mathcal{O}_B\rangle_c^4}{2\norm{\mathcal{O}_A}_2^4 \norm{\mathcal{O}_B}_2^4}, \quad 0 < \alpha \le 1
\end{align}
We now have three upper bounds for the connected correlation function: \eqref{eq:vNBound}, \eqref{eq:RenyiBound}, and \eqref{eq:RenyiBoundSchatten2}. Which one is the strongest depends on $\rho_{AB}$.

\bibliographystyle{JHEP}

\bibliography{main}
%~~~~~~~~~~~~~~~~~~~~~~~~~~~~~~~~~~~~~~~~~~~~~~~~~~~~~~~~~~~~~~~~~~~~~

\end{document}